\documentclass[pra,showpacs,twocolumn,superscriptaddress]{revtex4-1}

\usepackage{amssymb}
\usepackage{amsmath}
\usepackage{graphicx}
\usepackage{verbatim}
\usepackage{version}

\usepackage{color}
\definecolor{blue}{rgb}{0,0,1}
\definecolor{grey}{rgb}{0.6,0.6,0.6}

\def    \bse{\begin{subequations}}
\def    \ese{\end{subequations}}

\def \bew{\begin{widetext}\begin{equation}}
\def \eew{\end{equation}\end{widetext}}

\def \neff{\bar{n}_{\mathrm{eff}}}
\def \ntot{\bar{n}_{\mathrm{tot}}}

\def \neffone{\bar{n}_{\mathrm{eff},1}}
\def \ha{\hat{a}}
\def \had{\hat{a}^{\dag}}
\def \hd{\hat{d}}
\def \hdd{\hat{d}^{\dag}}
\def \g20{g^{(2)}(0)}
\def \gone{g^{(2)}_1(0)}
\def \gt{g^{(2)}(\tau)}
\def \aopt{\bar{\alpha}_{\mathrm{opt}}}
\def \ropt{r_{\mathrm{opt}}}
\def \ba{\bar{\alpha}}

\begin{document}

\title{Antibunching and unconventional photon blockade with Gaussian squeezed states}

\author{Marc-Antoine Lemonde}
\affiliation{Department of Physics, McGill University, 3600 rue University, Montreal, Quebec H3A 2T8, Canada}
\author{Nicolas Didier}
\affiliation{Department of Physics, McGill University, 3600 rue University, Montreal, Quebec H3A 2T8, Canada}
\affiliation{D\'epartment de Physique, Universit\'e de Sherbrooke, 2500 boulevard de l'Universit\'e, Sherbrooke, Qu\'ebec J1K 2R1, Canada}
\author{Aashish~A.~Clerk}
\affiliation{Department of Physics, McGill University, 3600 rue University, Montreal, Quebec H3A 2T8, Canada}

\date{\today}

\begin{abstract}
Photon antibunching is a quantum phenomenon typically observed in strongly nonlinear systems where photon blockade suppresses the probability for detecting two photons at the same time.
Antibunching has also been reported with Gaussian states, where optimized amplitude squeezing yields classically forbidden values of the intensity correlation, $\g20<1$.
As a consequence, observing antibunching is not necessarily a signature of photon-photon interactions. 
To clarify the significance of the intensity correlations, we derive a sufficient condition for deducing if a field is non-Gaussian based on a  $\g20$ measurement.
We then show that the Gaussian antibunching obtained with a degenerate parametric amplifier is close to the ideal case reached using dissipative squeezing protocols.
We finally shed light on the so-called \textit{unconventional photon blockade} effect predicted in a driven two-cavity setup with surprisingly weak Kerr nonlinearities, stressing that it is a particular realization of optimized Gaussian amplitude squeezing.
\end{abstract}


\maketitle



\section{Introduction}

Understanding, generating, and ultimately manipulating nonclassical states of light are fundamental goals of the field of quantum optics \cite{WallsMilburnBook}.  They have been pursued 
in a wide variety of physical systems ranging from atomic cavity QED systems and nonlinear optical media, to recent experiments with superconducting circuits; such states are also 
essential to many approaches to quantum information processing \cite{NielsenChuangBook}.  A common way to identify the quantumness of these states is to quantify their intensity fluctuations via the $\g20$ correlation function, defined as $\g20=\langle\colon\hat{I}^2\colon\rangle/\langle \hat{I} \rangle^2$ where $\hat{I}$ is the field intensity and colons indicate normal ordering.  While classical intensity fluctuations always obey $\g20 \geq 1$, quantum states can violate this bound; $\g20 < 1$ is hence often used as a criteria to identify nonclassical states.  

The standard mechanism for achieving $\g20 < 1$ is known as photon blockade~\cite{Imamoglu_PRL_1997,Birnbaum_Nature_2005,Bozyigit_Nature_2011}.   
A laser drives a nonlinear cavity in resonance with the $0 \rightarrow 1$ photon transition; however, because of the cavity's nonlinear spectrum, 
the laser cannot add another photon as the $1 \rightarrow 2$ photon transition is off resonant.  One thus obtains a strongly non-Gaussian state close to a single-photon Fock state, and strongly reduced intensity fluctuations.  
Observing photon blockade relies on the challenging task of having systems with nonlinear interactions that exceed the characteristic dissipation rate.
Similarly, phonon blockade arises in nonlinear mechanical resonators~\cite{Liu_PRA_2010, Didier_PRB_2011}.

\begin{figure}[tr]
  \centering
  \includegraphics[width=\columnwidth]{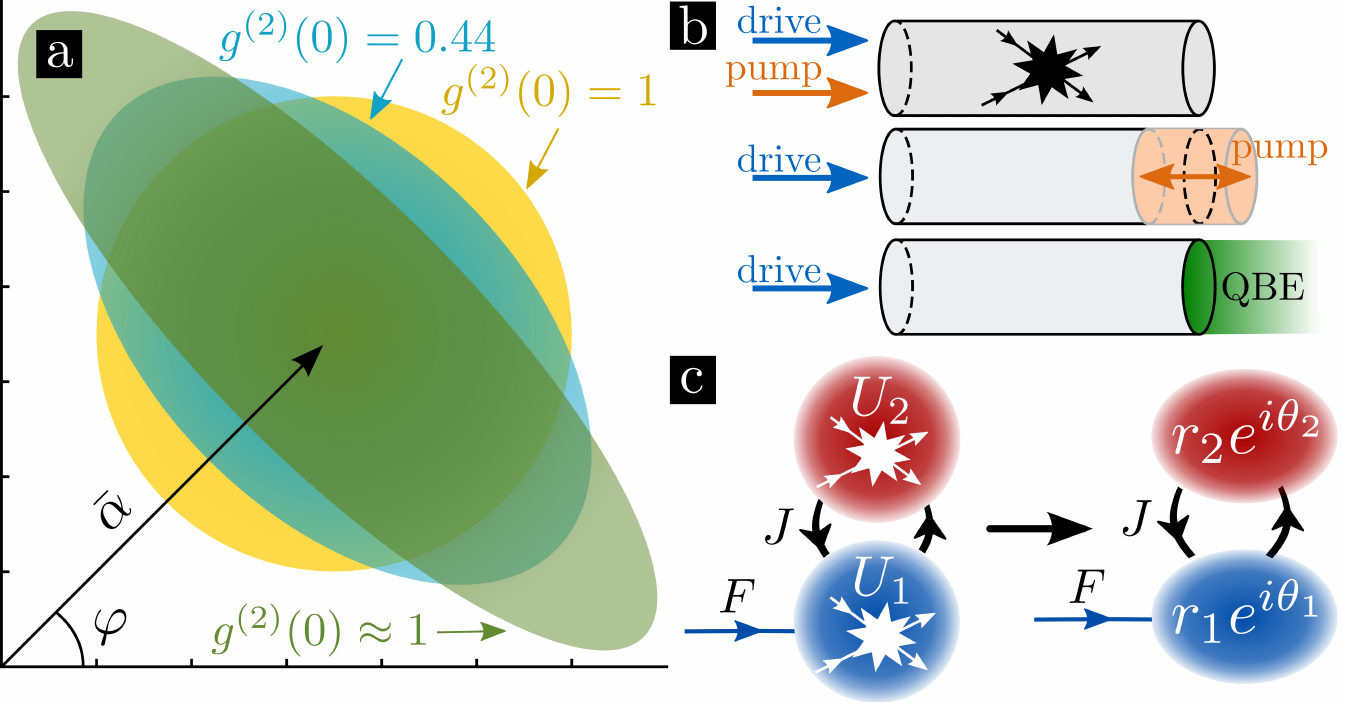}
  \caption{(a) Ideally amplitude-squeezed Gaussian states for a squeeze parameter $r=\ropt= 0.13$ (blue), $r = 0.3$ (green) and $r=0$ (yellow). 
(b) Realizations of a degenerate parametric amplifier with a pumped Kerr nonlinearity (top) and cavity frequency modulation (middle). 
Alternatively, highly pure intracavity squeezing can be generated using quantum bath engineering (QBE), where a cavity interacts with a structured reservoir (bottom).
(c) Two cavity setup where unconventional photon blockade is predicted. As pictured, the photon-photon interaction generates squeezing of the intracavity modes.}
 \label{Fig:Schema}
\end{figure}

Spurred both by recent studies in optomechanics and circuit QED~\cite{Rabl_PRL_2011, Bozyigit_Nature_2011, Kronwald_PRA_2013, Cohen_ArXiv_2014}, as well as by recent studies discussing a method for achieving $\g20<1$ with extremely weak nonlinearities~\cite{Liew_Savona_PRL_2010, Ferretti_PRA_2010, Bamba_PRA_2011, Bamba_Ciuti_APL_2011, Ferretti_Savona_NJP_2013, Liew_Savona_NJP_2013, Flayac_Savona_PRA_2013, Xu_Li_JPB_2013}, we revisit in this work a somewhat under-appreciated fact:   
the $\g20<1$ condition for nonclassicality can be achieved without photon blockade, by simply using optimized amplitude-squeezed Gaussian states.  The basic mechanism is depicted in Fig~\ref{Fig:Schema}(a),  and was discussed in several previous works~\cite{Stoler_PRL_1974, Mahran_PRA_1986, Koashi_PRL_1993, Lu_PRL_2001,Grosse_PRL_2007}.  The upshot is that states with nonclassical intensity fluctuations can be generated using 
purely \emph{linear} bosonic systems (i.e.~described by a quadratic Hamiltonian), without the need of any spectral nonlinearity, and without any negativity in the Wigner function of the state.  It also leads to the conclusion that some Gaussian states are more quantum than others (in that not all violate the classical bound on $\g20$).

\begin{figure*}[t]
  \centering
  \includegraphics[width=\textwidth]{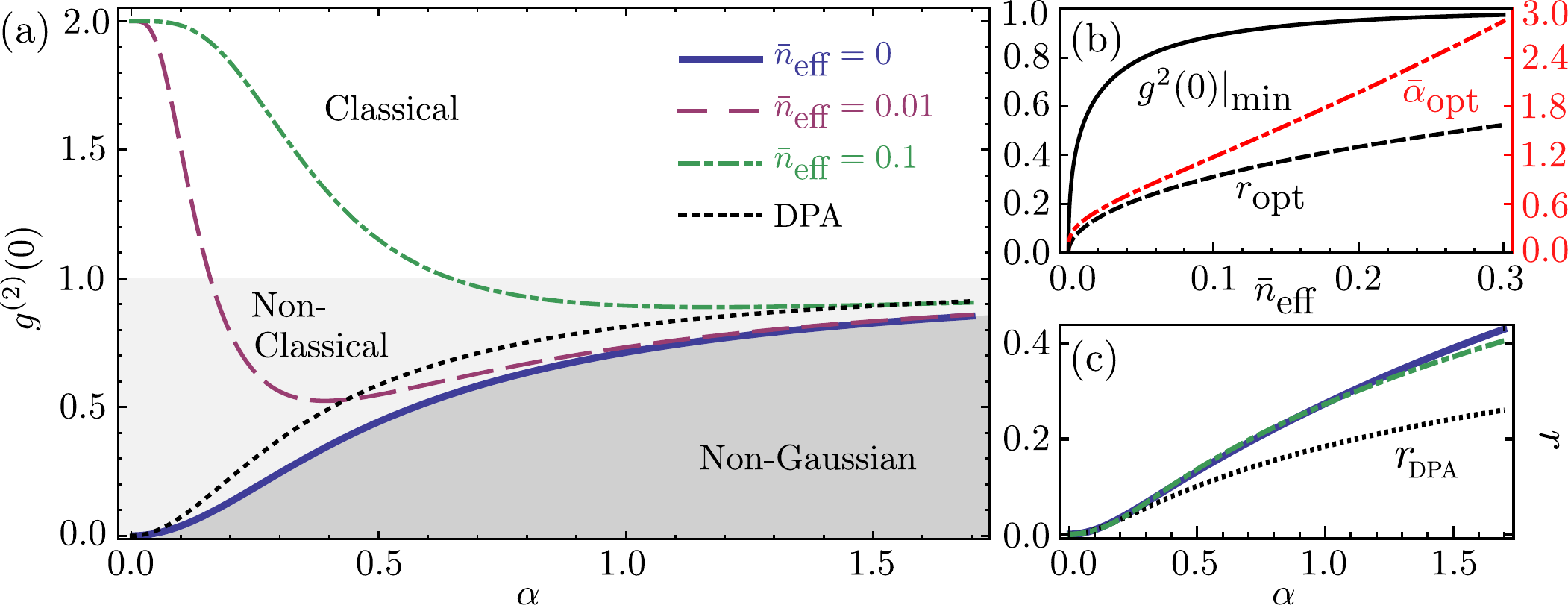}
  \caption{(a) Equal-time intensity correlation function $\g20$ for a Gaussian state as a function of the displacement $\ba$ for different values of the state purity (as quantified by $\neff$). For all curves the squeeze parameter has been set to its optimal value, $r = \ropt[\ba,\neff]$; $\ropt$ is plotted in panel (c) as a function of $\ba$.
The dotted black line corresponds to the degenerate parametric amplifier (DPA) (see following section) where $\neff=\sinh^2r_\mathrm{DPA}$.
The curve corresponding to $\neff=0$ (solid blue) sets the minimum value of $\g20$ possible for a Gaussian state with $|\langle \hat{a} \rangle| = \ba$; any values lying in the darkest shaded region necessarily corresponds to non-Gaussian states. For finite $\neff$, $\g20 = 2$ for $\alpha = 0$ ($r=0$ also), as it should for a thermal state. 
Panel (b) presents the minimal $\g20 \vert_{\mathrm{min}}$ (black full line) that can be achieved with a Gaussian state having a fixed value of $\neff$. For these curves, $\alpha = \aopt$ (red dot-dashed line) and $r = \ropt$ (black dashed line) [cf.~Eq.~(\ref{Eq:ropt})].} \label{Fig:g20General}
\end{figure*}

In this work, we start by characterizing the equal-time intensity correlation function of the most general single-mode Gaussian state (i.e.~a displaced, squeezed thermal state), identifying the full parameter regime where $\g20 < 1$ (see also Ref.~\cite{Grosse_PRL_2007}).  We show that for a given average field amplitude, there is a minimum possible value of $\g20$ consistent with a Gaussian state.  Often, a finding of $\g20 < 1$ is used implicitly (and incorrectly) as evidence of a non-Gaussian state.  The results presented here allow one to simply identify when a measurement of $\g20 < 1$ necessarily signals the existence of a non-Gaussian state (see dark shaded region in Fig.~\ref{Fig:g20General}).

We also discuss how this Gaussian-state $\g20$ suppression can be realized using one of two simple and generic cavity-based setups: either via a degenerate parametric amplifier (DPA), as
has been studied earlier \cite{Koashi_PRL_1993, Lu_PRL_2001, Grosse_PRL_2007}, or via dissipative squeezing interactions~\cite{Cirac_Zoller_PRL_1993, Rabl_PRB_2004, Parkins_PRL_2006, DallaTorre_PRL_2013, Tan_PRA_2013, Didier_PRA_2014, Kronwald_PRA_2013_DissSqueezing}.
While dissipative squeezing has the virtue of being able to produce pure squeezed intracavity states, we find that in terms of $\g20$ suppression, it only gives a marginal improvement over a DPA.  Further, we show that the DPA exhibits a kind of optimality:  for a given level of state impurity, the amount of squeezing produced is exactly what is needed to allow a maximal $\g20$ suppression.
We also show how these two generic approaches lead to photon antibunching and non-monotonic behavior of the two-time intensity correlation function $\gt$. These are additional nonclassical features characterized respectively by the conditions $\gt > \g20$ and $\vert \gt - 1 \vert > \vert \g20 - 1 \vert$~\cite{Rice_IEEE_1988}. 

Finally, we use this Gaussian-state $\g20$ suppression mechanism to help clarify a series of recent studies~\cite{Liew_Savona_PRL_2010, Ferretti_PRA_2010, Bamba_PRA_2011, Bamba_Ciuti_APL_2011, Ferretti_Savona_NJP_2013, Liew_Savona_NJP_2013, Flayac_Savona_PRA_2013, Xu_Li_JPB_2013} of a novel driven nonlinear two-cavity setup where $\g20 <1$ is predicted despite having nonlinearities much weaker than all dissipative rates.  Nonlinearity was suggested to be the key ingredient~\cite{Bamba_PRA_2011} behind this so-called \textit{unconventional photon blockade} (UPB)~\cite{Flayac_Savona_PRA_2013}. Here we show that the UPB is more simply understood as yet another realization of the optimally-squeezed Gaussian state mechanism for $\g20$ suppression.

\section{Intensity fluctuations of general Gaussian states} \label{Sec:GaussianStares}

The simplest measure of the intensity fluctuations of a light field (as measured by a photomultiplier) is the $\gt$ correlation function,
\begin{equation} \label{Eq:g2definition}
	\gt \equiv \frac{\left\langle \had(0) \had(\tau) \ha(\tau) \ha(0)\right\rangle}
	{\left\langle \had(0) \ha(0) \right\rangle \left\langle \had(\tau) \ha(\tau) \right\rangle },
\end{equation}
where $\ha$ is the photon annihilation operator.
$\gt$ is proportional to the conditional probability for detecting a second photon at time $t=\tau$, given that a photon was also detected earlier at $t = 0$; 
it can be measured using a Hanbury Brown and Twiss type experiment~\cite{HBT_Nature_1956,Bromberg_Nature_2010}. 
For classical (commuting) fields, it directly follows from the Cauchy-Schwarz inequality that $\g20 \geqslant 1$, as well as $\gt \leqslant \g20$~\cite{GerryKnightBook}. In contrast, a quantum field prepared in an appropriate state can violate one or both of these classical bounds; such states are generally termed ``nonclassical". Our first goal here will be to remind the reader that a properly optimized Gaussian state can lead to such nonclassical signatures \cite{Stoler_PRL_1974, Mahran_PRA_1986, Koashi_PRL_1993, Lu_PRL_2001,Grosse_PRL_2007}.

The most general single-mode Gaussian state, i.e.~a displaced squeezed thermal state, is described by the density matrix
\begin{equation} \label{Eq:GaussianState}
	\hat{\rho}_{\alpha, \xi, \neff} \equiv \hat{D}(\alpha) \hat{\rho}_{\neff, \xi} \hat{D}^{\dag}(\alpha) = \hat{D}(\alpha)\hat{S}(\xi) \hat{\rho}_{\neff} \hat{S}^{\dag}(\xi)\hat{D}^{\dag}(\alpha). 
\end{equation}
Here, $\hat{D}(\alpha) = \exp[\alpha \had - \alpha^* \ha]$ and $\hat{S}(\xi) = \exp[\frac{1}{2}( \xi^* \ha^2 - \xi \hat{a}^{\dag 2})]$ are respectively the displacement and squeezing operators~\cite{GerryKnightBook}, with $\alpha = \ba e^{i\varphi}$ and $\xi = r e^{i\theta}$ ($\ba > 0, r > 0$).  $\hat{\rho}_{\neff}$ is the density matrix of a thermal state with population $\neff$; $\mathrm{Tr}\left[ \hat{\rho}_{\neff} \had\ha \right]  = \neff$. 
The purity $P$ of the density matrix in Eq.~\eqref{Eq:GaussianState} is set by $\neff$ according to the relation
\begin{equation}
P\equiv\mathrm{Tr}[\hat{\rho}_{\alpha, \xi, \neff}^2]=\frac{1}{1+2\neff}. \label{Eq:P}
\end{equation}

As one might expect, to minimize intensity fluctuations it is always optimal to squeeze the amplitude quadrature, i.e.~choose $\theta = 2 \varphi$.  In this case, we find
\begin{subequations}
\begin{align}
	\g20 & = 1+\frac{2 \ba^2 (n - s)+s^2 + n^2}{\left(\ba^2+n\right)^{2}}, \label{Eq:g20Gaussian} \\
	\ntot & \equiv \langle \hat{a}^\dagger \hat{a} \rangle =  \ba^2 + n, \label{Eq:ntot} 
\end{align}
\end{subequations}
with
\begin{subequations} \label{Eq:ns}
\begin{align}
	n & \equiv \mathrm{Tr}\left[ \hat{\rho}_{\neff, \xi} \had\ha \right] = (\neff + \tfrac{1}{2})\cosh2r-\tfrac{1}{2}, \label{Eq:n}\\
	s & \equiv |\mathrm{Tr}\left[ \hat{\rho}_{\neff, \xi} \ha\ha \right]|  = (\neff + \tfrac{1}{2})\sinh2r.
\end{align}
\end{subequations}
Using Eq.~(\ref{Eq:g20Gaussian}), it is now straightforward to find conditions on the displacement, squeezing and effective temperature of our Gaussian state that reduce $\g20$ below 1. Note that for arbitrary angles, $(n-s) \rightarrow n-s\cos(\theta - 2\varphi)$ in Eq.~\eqref{Eq:g20Gaussian}.

We first investigate the ideal (and optimal) case where we have a pure state, i.e.~$\neff = 0$.  Eq.~(\ref{Eq:g20Gaussian}) then simplifies to:
\begin{equation}
	\g20 = 1+\frac{\cosh2r}{\ba^2 + \sinh^2r} - \frac{\ba^2(1 + \sinh2r)}{(\ba^2 +\sinh^2r)^2}. \label{Eq:g20neff0}
\end{equation}
This simple expression already reveals some surprises.  For no squeezing (i.e.~$r=0$), we recover a coherent state and $\g20 = 1$.  One might have expected that $\g20$ would decrease monotonically if we now start to increase $r$ (i.e.~the greater the squeezing, the smaller the intensity fluctuations).  This is clearly incorrect:  
Eq.~(\ref{Eq:g20neff0}) yields $\g20 \rightarrow 3$ as $r \rightarrow \infty$, as was experimentally observed in  Ref.~\cite{Grosse_PRL_2007}.
Eq.~(\ref{Eq:g20neff0}) instead reveals that for a fixed displacement $\ba$, $\g20$ has a minimum as a function of $r$; we denote this optimal value $\ropt[\ba,\neff=0]$.
If $r$ is tuned to $\ropt[\ba,\neff=0]$, we find that the resulting $\g20$ is always less than one (and hence nonclassical), no mater how large the displacement $\ba$.  
However, as $\ba \rightarrow \infty$, the optimized $\g20$ approaches the classical value of $1$ from below as $\g20 \rightarrow 1 - 1/\ba^2$.
The full behavior of $\g20$ versus $\ba$ for this optimally-squeezed Gaussian state (for $\neff = 0$) is shown in Fig.~\ref{Fig:g20General}. 

For further insight, it is useful to consider the limit of small displacements, $\ba \ll 1$.  The optimal squeezing and the corresponding $\g20$ are given by
\begin{equation}
	\ropt\left[ \ba, \neff=0 \right] \approx \ba^2 + \mathcal{O}[\ba^4]  
	\Rightarrow \g20 \approx 4\ba^2 + \mathcal{O}[\ba^4]. \label{Eq:ropt_alpha_neff}
\end{equation}
Thus, one can make $\g20$ as small as one likes by simply taking a small enough displacement $\ba$ and always picking the optimal (small) amount of squeezing.  

The optimal parameter values in the small $\ba$ limit are easily understood.  To leading non-vanishing order in $\ba$ and $r$, the probability for having two photons in our state is 
$\left| \langle2|D(\alpha)S(\xi)|0\rangle \right|^2   \approx   \left( \ba^2-r \right)^2/2$.  The two terms here indicate the two ways of getting two photons:  either via the squeeze operator, or via the displacement operator.  The optimal squeezing condition thus simply corresponds to these two mechanisms interfering destructively~\cite{Lu_PRL_2001}. Note that unlike photon blockade, we are {\it not} using the nonlinearity of the spectrum to suppress the two-photon population, but rather the interference between a coherent displacement and a squeeze operation. 

Returning to the case of a general $\ba$, we see that for a given average amplitude $|\langle \hat{a} \rangle| = \ba$, there is a minimum possible $\g20$ achievable with a Gaussian state; if one obtains a lower $\g20$, this then necessarily implies that the state is non-Gaussian.   For example, for $\ba = 1$ the smallest $\g20$ achievable with a Gaussian state is $\g20 \approx 0.71$, which is achieved when the squeeze parameter $r \approx 0.28$. 
This general bound on the minimal $\g20$ for a Gaussian state is shown in Fig.~\ref{Fig:g20General}:  the dark shaded region indicates regimes where the intensity fluctuations are both too small to be explained classically, or be explained by a Gaussian state.  We stress that for states with $\ba \rightarrow 0$, an arbitrarily small value of $\g20$ is possible with a Gaussian state, and hence in this case a $\g20$ measurement cannot be used to conclusively prove the existence of a non-Gaussian state.

We also show in Fig.~\ref{Fig:g20General} that $\g20 < 1$ is possible with a Gaussian state even if it fails to be pure, i.e.~if $\neff$ is non-zero in Eq.~(\ref{Eq:g20Gaussian}). 
Suppose we only consider Gaussian states which have a fixed effective thermal number $\neff$:  if we optimize both the state displacement $\ba$ and the squeezing magnitude $r$ for such states, 
how small can we make $\g20$?  The optimal amount of squeezing $\ropt$ is, in this case, determined by
\begin{align}
	\sinh^2\ropt=\neff. \label{Eq:ropt}
\end{align}
From Eq.~(\ref{Eq:n}), one sees that $\sinh^2r$ is the mean number of excitations in a vacuum squeezed state having squeeze parameter $r$; for $\ropt$, the contributions to the mean number of excitations due to squeezing and due to thermal fluctuations are equal.

The corresponding full expression for the optimal displacement $\ba = \aopt$ is given in Appendix~\ref{App:Opt_g2} (Eq.~\eqref{Eq:alphaOpt_rneff}); 
for the most interesting case of a small thermal population $\neff \ll 1$, 
it is approximately given by $\aopt \sim \neff^{1/4}$. These choices lead to a minimal $\g20 \vert_{\mathrm{min}} \approx 8\sqrt{\neff}$ (see inset of Fig.~\ref{Fig:g20General} and Appendix~\ref{App:Opt_g2} for general $\neff$). Thus, for an impure state, one cannot suppress $\g20$ arbitrarily, even if one takes an arbitrarily small displacement $\ba$.  It is also interesting to note that the optimal relation between $\neff$ and $r$ given in Eq.~(\ref{Eq:ropt}) is exactly satisfied by a DPA; we will see this explicitly below.

\section{Antibunching using coherent and dissipative squeezing interactions}

Having described the suppression of $\g20$ below 1 for optimized amplitude-squeezed Gaussian states, we now discuss a simple system which can realize this physics via coherent squeezing interactions, namely a cavity-based DPA~\cite{Koashi_PRL_1993,Lu_PRL_2001, Grosse_PRL_2007}. We also compare it to schemes that achieve squeezing via dissipative interactions~\cite{Cirac_Zoller_PRL_1993, Rabl_PRB_2004, Parkins_PRL_2006, DallaTorre_PRL_2013, Tan_PRA_2013, Didier_PRA_2014, Kronwald_PRA_2013_DissSqueezing}.  Such schemes have recently garnered interest in both the circuit QED ~\cite{Didier_PRA_2014} and 
optomechanics communities \cite{Kronwald_PRA_2013_DissSqueezing}, and have the virtue that they can in principle generate pure intracavity squeezing. 
We focus here on the intracavity $\gt$; this is of direct relevance in experiments in optomechanics~\cite{Cohen_ArXiv_2014}, and as we show at the end of this section, it is simply related to the correlations of the output field.
We then recover the fact that the DPA exhibits true photon antibunching~\cite{Koashi_PRL_1993,Lu_PRL_2001, Grosse_PRL_2007} and nonclassical non-monotonic behaviors characterized by $\vert \gt - 1 \vert > \vert \g20 - 1 \vert$~\cite{Rice_IEEE_1988, Lu_PRL_2001}.
We stress that in our case, the temporal evolution of the output correlations is only due to the intracavity dynamics; in contrast, the time-dependence of $\gt$ calculated in Ref.~\cite{Grosse_PRL_2007} only reflected the bandwidth of the chosen filter. 

The general Hamiltonian of the DPA reads, in the interaction picture ($\hbar=1$),
\begin{equation}
\hat{H} =-\tfrac{i}{2}\lambda[\hat{a}^{\dag2}e^{i\theta}-\hat{a}^2e^{-i\theta}]
+[\epsilon\had+\epsilon^*\ha]+\hat{H}_\kappa,
\label{HDPA}
\end{equation}
where the strength of the squeezing interaction $\lambda$ and angle $\theta$ are controlled by the setup characteristics.
The Hamiltonian $\hat{H}_\kappa$ describes the coupling to a single input/output waveguide; this coupling is characterized by the damping rate $\kappa$ and is treated with a standard input-output approach~\cite{Clerk_RMP_2010, GardinerZollerBook} (see also Appendix~\ref{App:DPA}).
We assume that the phase of the parametric coupling is always tuned to satisfy $\theta=2\arg{\langle\hat{a}\rangle}$, i.e.~to get an amplitude-squeezed intracavity state.
The system is stable for $\lambda \leq \kappa/2$, and the steady state inside the cavity is a displaced squeezed thermal state with the parameters
$\tanh 2r=2\lambda/\kappa$ and 
$
\neff=\sinh^2r.
$
Remarkably, the latter relation is identical to Eq.~\eqref{Eq:ropt}, (i.e.~the optimal amount of squeezing given a fixed effective thermal number).

For a fixed value of parametric coupling $\lambda$ in the DPA, the optimal displacement needed to minimize the intracavity $\g20$ is given by $\aopt^2=\kappa\lambda/(\kappa-2\lambda)^2$; this can be obtained by tuning the drive strength $\epsilon$.  Alternatively, one could imagine minimizing $\g20$ for a fixed value of $\ba$ (by tuning both $\lambda$ and $\epsilon$ simultaneously).  The resulting minimal $\g20$ versus $\ba$ is plotted in Fig.~\ref{Fig:g20General}
and is extremely close to the minimal value allowed for an arbitrary Gaussian state with the same $\ba$.  
While $\g20$ is minimal for infinitesimally small displacement and squeezing, $\g20<1$ can be obtained even for displacements at the single to few photon level.

\begin{figure}[t]
\includegraphics[width=\columnwidth]{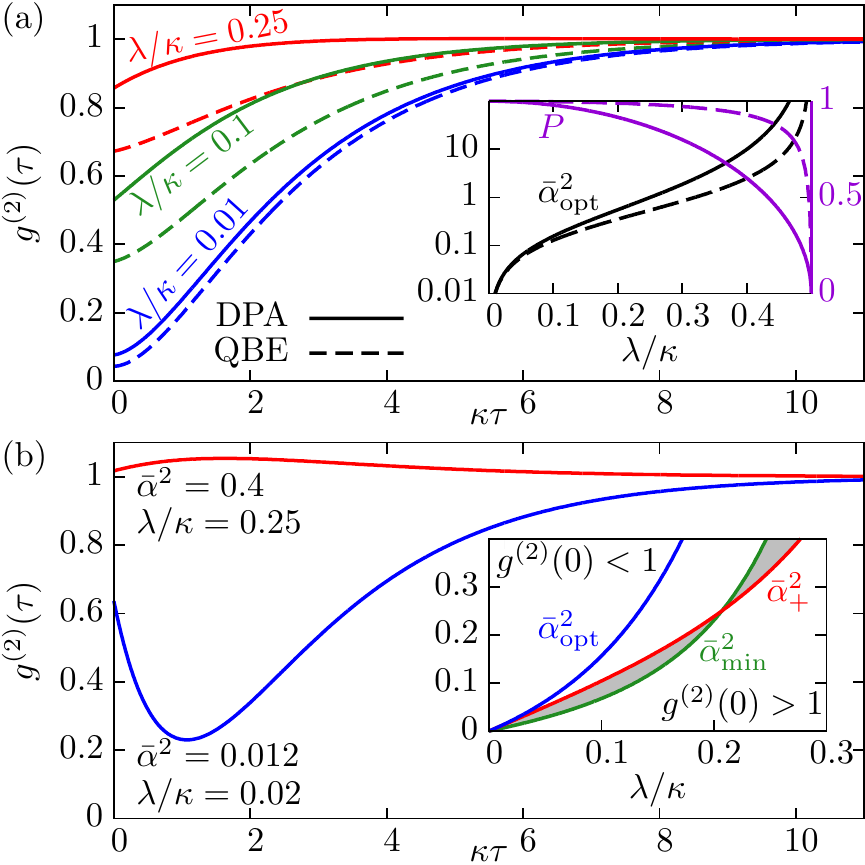}
\caption{Antibunching in the DPA versus QBE. 
(a) Two-time intensity correlation $\gt$ 
versus $\tau$.  Solid curves are for the DPA system, with different curves corresponding to different values of the strength of the squeezing interaction $\lambda$; for each curve, the coherent displacement $\ba = \aopt$.  The dashed lines are for the QBE system.
The intracavity squeeze parameter $r$ is the same in the two cases and set by the value of $\lambda/\kappa$.
For QBE $\Gamma_\mathrm{QBE}=0.9\,\kappa$.
The optimal displacement $\aopt$ and the purity $P$ of the intracavity state (cf.~Eq.~\eqref{Eq:P}) is plotted in the inset as a function of $\lambda/\kappa$.
For $\Gamma_\mathrm{QBE}\gg\kappa_{\rm ext}$, QBE generates a much better purity than the DPA.
(b) Nonclassical non-monotonic behavior of $\gt$ for a DPA; parameters for both curves are indicated in the plot.
Inset: $\ba_{\mathrm{min}}$ (green line) is the minimum value of $\ba$ for the given $\lambda / \kappa$ for which $\g20 < 1$, while $\ba_+$ (red line) represents the minimum value of $\ba$ for which we have true antibunching (i.e.~the slope of $\gt$ at $\tau=0$ is positive). Values of $\ba$ between these curves (shaded region) will yield $\gt$ functions which are non-monotonic.}
\label{figg2t}
\end{figure}

To show furthermore that the field is antibunched, we calculate the intracavity two-time intensity correlation $\gt$,
\begin{equation}
\gt = 1+\frac{2\ba^2[n(\tau)-s(\tau)]+n^2(\tau)+s^2(\tau)}{[\ba^2+n(0)]^{2}},
\end{equation}
with $\tau\geq0$ and the two-time correlations 
$n(\tau)\equiv\langle\hat{d}^\dag(\tau)\hat{d}^\dag(0)\rangle$
and
$s(\tau)\equiv|\langle\hat{d}(\tau)\hat{d}^\dag(0)\rangle|$ (with $\hd \equiv \ha - \langle \ha \rangle$) given by
\begin{subequations}
\begin{align}
n(\tau)&=\frac{\lambda\,e^{-\kappa\tau/2}}{\kappa^2-4\lambda^2}\left[2\lambda\cosh(\lambda\tau)+\kappa\sinh(\lambda\tau)\right],\\
s(\tau)&=\frac{\lambda\,e^{-\kappa\tau/2}}{\kappa^2-4\lambda^2}\left[2\lambda\sinh(\lambda\tau)+\kappa\cosh(\lambda\tau)\right].
\end{align}
\end{subequations}
The time-dependent intensity correlations $\gt$ are plotted in Fig.~\ref{figg2t} for different values of $\lambda/\kappa$.
The correlations start well below unity and increase to reach 1 at long times.
This behavior is the signature of photon antibunching, which has been measured experimentally in Refs.~\cite{Koashi_PRL_1993,Lu_PRL_2001, Grosse_PRL_2007}.
More precisely, the condition to have a positive slope at $\tau=0$, $\dot{g}^{(2)}(0)>0$, corresponds to $\ba^2>\kappa\lambda/(\kappa^2-4\lambda^2) \equiv \ba_+^2$.
This condition for antibunching is always satisfied for $\ba=\aopt$. Moreover, in Fig.~\ref{figg2t}(b), we show an additional nonclassical behavior that can be exhibited by $\gt$, where $\vert \gt - 1 \vert > \vert \g20 - 1 \vert$. In order to observe this behavior, one has either to tune $\ba$ such that $\g20 < 1$ and $\dot{g}^{(2)}(0)<0$ or such that $\g20 > 1$ and $\dot{g}^{(2)}(0)>0$. These two conditions are represented by the shaded region in Fig.~\ref{figg2t}(b). The former case has been observed in~\cite{Lu_PRL_2001}. 

The generation of intracavity vacuum squeezed states is also possible using QBE approaches; unlike the DPA where one is using a coherent (Hamiltonian) squeezing interaction, one is now making use of dissipative squeezing interactions.  There are a variety of methods for achieving this kind of interaction, e.g.~by modulating the cavity damping rate as a function of time~\cite{Didier_PRA_2014} or via two-tone driving~\cite{Kronwald_PRA_2013_DissSqueezing}.  The ideal versions of such scheme are equivalent to having effectively coupled the cavity to a squeezed reservoir (as could be realized directly by driving the cavity with vacuum squeezed light~\cite{Murch_Nature_2013}).

We wish to compare the ability of such dissipative-squeezing approaches to generate states with $\g20 < 1$ against the coherent-squeezing approach (i.e.~using a DPA).
We model the dissipative squeezing interaction by taking the cavity to be coupled to a Markovian squeezed reservoir (in addition to the main port used to drive the cavity and extract an output field).
This additional reservoir is characterized by a squeeze parameter $r_{\rm QBE}$ and an additional cavity damping rate $\Gamma_\mathrm{QBE}$.  
The damping rate $\Gamma_\mathrm{QBE}$ adds with the external damping rate $\kappa_{\rm ext}$ due to the coupling to the input/output waveguide to give the total cavity damping $\kappa_\mathrm{QBE}$.  
The quantum master equation $\dot{\hat{\rho}}=\hat{L}\hat{\rho}$ of the density matrix $\hat{\rho}$ of the system induced by the two reservoirs is governed by the following Lindbladian,
\begin{equation}
\hat{L}\hat{\rho}=\Gamma_\mathrm{QBE}[\hat{b}\hat{\rho}\hat{b}^\dag-\tfrac{1}{2}\{\hat{b}^\dag\hat{b},\hat{\rho}\}]
+\kappa_\mathrm{ext}[\hat{a}\hat{\rho}\hat{a}^\dag-\tfrac{1}{2}\{\hat{a}^\dag\hat{a},\hat{\rho}\}].
\label{LindbladQBE}
\end{equation}
The Lindbladian has two contributions.
The first term describes dissipative squeezing~\cite{Didier_PRA_2014} and cools the Bogoliubov mode
$\hat{b}=\cosh r_\mathrm{QBE}\,\hat{a}+e^{i\theta/2}\sinh r_\mathrm{QBE}\,\hat{a}^\dag$ 
to its vacuum; this corresponds to the vacuum squeezed state $|r_\mathrm{QBE}e^{i\theta}\rangle$ for the intracavity field $\hat{a}$.
The second term cools the intracavity field to vacuum, thereby altering the squeezing purity.
The purity tends to unity for $\Gamma_\mathrm{QBE}\gg\kappa_\mathrm{ext}$, as can be seen in Fig.~\ref{figg2t}.
Further details are given in Appendix~\ref{App:SqueezedEnv}.

The comparison of the dissipative squeezing setup against the DPA is shown in Fig.~\ref{figg2t}, where we plot the two-time intensity correlation $\gt$ for both approaches.  
To facilitate comparison, we use parameters that ensure that the schemes produce equivalent amounts of intracavity squeezing (see Appendix~\ref{App:QBEcomp}).
We also pick $\Gamma_\mathrm{QBE} = 9 \kappa_{\rm ext} =  0.9 \kappa_{\rm QBE} $ to ensure that the QBE scheme generates a highly pure squeezed state, and that the total cavity damping rate is the same in both approaches, $\kappa_\mathrm{QBE}=\kappa$.  Finally, in both schemes the coherent drive is chosen so as to yield an optimal $\ba$, i.e. a value which minimizes $\g20$.
Fig.~\ref{figg2t} shows that despite the additional purity achieved using the dissipative scheme, it does not perform significantly better except at the largest values of $\lambda / \kappa$ (which corresponds to the largest values of $r$ and of $\ba$).
The purity and the optimal $\g20$ of the two setups are equal when the two damping rates coincide, $\Gamma_\mathrm{QBE} = \kappa_{\rm ext}$.

Finally, while we have discussed intracavity fields here, our results are easily extended to the $\gt$ function of the output light field leaving the coupled waveguide.  For a single sided cavity case we focus on, the expressions for the output field $g^{(2)}_\mathrm{out}(\tau)$ are identical to those for the intracavity $\gt$, except that one needs to replace $\alpha$ with $\alpha + \alpha_\mathrm{in}/\sqrt{\kappa}$ (see Appendix~\ref{App:OutputField}).

\section{Two-cavity unconventional photon blockade}

Having discussed in detail the fact that optimally-squeezed Gaussian states can lead to the nonclassical regime $\g20 < 1$, we revisit the system introduced by Liew and Savona~\cite{Liew_Savona_PRL_2010} (subsequently studied in Refs.~\cite{Ferretti_PRA_2010, Bamba_PRA_2011, Bamba_Ciuti_APL_2011, Ferretti_Savona_NJP_2013, Liew_Savona_NJP_2013, Flayac_Savona_PRA_2013}). This work predicts $\g20 < 1$ in a two-cavity system having extremely weak Kerr nonlinearities (i.e.~onsite photon-photon interactions). We show here that the main effect in the Liew and Savona system can be explained entirely using Gaussian states. The only role of the Kerr interaction in the two-cavity system is thus to provide an effective (quadratic) squeezing term in the Hamiltonian; the extremely weak nonlinearity of system's spectrum plays no role.  Moreover, the interpretation of the UPB as a result of interference between different paths leading to the two-photons state, as proposed in \cite{Bamba_PRA_2011}, is exactly the same interference condition that leads to maximal suppression of $\g20$ as discussed below Eq.~(\ref{Eq:ropt_alpha_neff}) in Sec.~\ref{Sec:GaussianStares} and in Ref.~\cite{Lu_PRL_2001}.
The Liew-Savona system is thus a particular realization of the general Gaussian-state mechanism for the suppression of $\g20$ discussed above, albeit a more complicated one than the single-cavity DPA.

The Liew-Savona system is composed of two bosonic modes, e.g.~optical modes in two separated cavities (as pictured in Fig.~\ref{Fig:Schema}). In each cavity, Kerr-type photon-photon interaction takes place with interaction strength $U_k$ ($k = 1,2$). The two cavities are linearly coupled together by a hopping term (rate $J$) and a weak drive $F$ (at frequency $\omega_d$) is applied to the first cavity only. The corresponding Hamiltonian, in the frame rotating at the drive frequency, is given by
\begin{align}
\hat{H} = \sum_{k=1}^2 & [ -\Delta_k\had_k \ha_k + U_k \had_k\had_k\ha_k\ha_k ] \nonumber \\
& + J[ \had_1\ha_2 + \had_2\ha_1 ] + F[ \had_1 + \ha_1 ]. \label{Eq:TwoCavHamiltonian}
\end{align}
Here, $\ha_k$ is the annihilation operator of mode $k$ with detuning $\Delta_k$ and $\hat{H}_{\kappa}$ describes the coupling to the environment characterized by the damping rates $\kappa_k$ of each cavities. We consider $J$ and $F$ to be real and positive without loss of generality. From Eq.~(\ref{Eq:TwoCavHamiltonian}), we derive the Heisenberg-Langevin equations of motion for both cavities using input-output formalism~\cite{Clerk_RMP_2010, GardinerZollerBook}.  One can solve these equations approximately by linearizing them about the classical steady-state solution (see Appendix~\ref{App:TwoCav}).  This approximation is equivalent to treating the interactions in Eq.~(\ref{Eq:TwoCavHamiltonian}) at a mean-field level:  we are thus approximating the system with a quadratic Hamiltonian, and precluding any effects associated with non-Gaussian states and spectral nonlinearity.  In this approximation, the resulting steady-state intracavity field of cavity 1 (where the interesting physics is predicted) is Gaussian, and thus has the form  
$\hat{\rho}_{\alpha_1, \xi_1, \neffone}$ given in Eqs.~(\ref{Eq:GaussianState}),  
where we parameterize $\xi_1 = r_1 e^{i\theta_1}$ and $\alpha_1 = \ba_1 e^{i\varphi_1}$. 

\begin{figure}[t]
  \centering
  \includegraphics[width=\columnwidth]{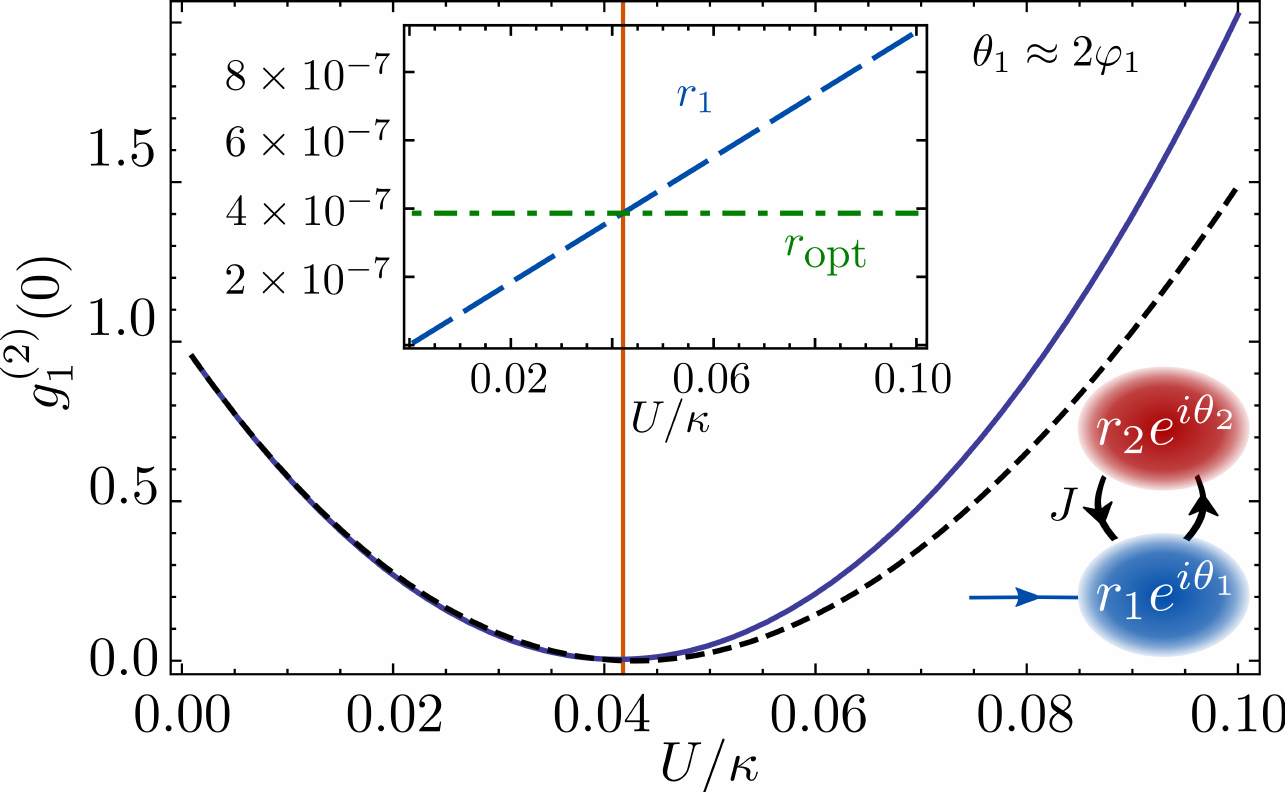}
  \caption{$\g20$ of the field inside cavity-1 of the two-cavity setup described in the main text as a function of the Kerr interaction strength $U$. 
The full line is the solution of the linearized Langevin equations and the dashed line is the solution of the quantum master equation (see Appendix~\ref{App:QME}).
In the inset, we compare the corresponding squeeze parameter $r_1$ (blue dashed line) to the optimal value $\ropt[\ba_1, \neffone=0]$ (green dot-dashed line) and we see that full suppression of $\gone$ corresponds to the case $r_1 = \ropt[\ba_1, \neffone=\sinh^2r_1]$. For these curves, we use the same parameters as in~\cite{Bamba_PRA_2011}, namely $\kappa_1 = \kappa_2 = \kappa$, $U_1 = U_2 = U$, $\Delta_1 = \Delta_2 = -0.275\kappa$, $F = 0.01\kappa$ and $J=3\kappa$. For such values $\neffone \sim 10^{-13}$ and the total photon number inside the cavity is $\bar{n}_{\mathrm{tot},1} \sim 10^{-7}$ [cf.~Eq.(\ref{Eq:ntot})].}
\label{Fig:TwoCavities}
\end{figure}

In Fig.~\ref{Fig:TwoCavities}, we show how the cavity-1 squeeze parameter $r_1$ and $\gone$ behave as a function of the interaction strength $U = U_1 = U_2$ for the same choice of parameters $J$, $\kappa_k$ and $\Delta_k$ as in Ref.~\cite{Bamba_PRA_2011}.  Results from the approximate linearized dynamics are shown, compared against results of a numerical solution of the full quantum master equation  (see Appendix~\ref{App:QME}).  The two approaches are in excellent agreement for the range of $U$ yielding a minimal $\g20$; in particular, the linearized dynamics accurately describe the results obtained in Ref.~\cite{Bamba_PRA_2011}.  This implies that the maximal $\g20$ suppression seen here is completely due to an optimally-squeezed Gaussian state in cavity 1.  

The connection to optimized Gaussian squeezing can be made even more precise.  From the solutions of the linear equations of motion, one can show that both
the cavity-1 squeeze angle $\theta_1$ and cavity-1 average amplitude $\ba_1e^{i\varphi_1}$ are
(to an excellent approximation) constant over the range of $U$ considered in Fig.~\ref{Fig:TwoCavities}. For the set of parameters used, they correspond almost precisely to amplitude squeezing ($\theta_1 - 2\varphi_1 \approx 0.065$). Also, for the whole range of $U$ in Fig.~\ref{Fig:TwoCavities}, the relation between $r_1$ and $\neffone$ is the same as a DPA (i.e.~$\neffone = \sinh^2r_1$). Thus, the only parameter determining the cavity-1 Gaussian state that varies with $U$ in Fig.~\ref{Fig:TwoCavities} is $r_1$, the magnitude of the cavity-1 squeezing.  As shown in the inset of Fig.~\ref{Fig:TwoCavities}, maximal suppression of $\gone$ exactly corresponds to the case where $r_1 = \ropt[\ba_1,\neffone=\sinh^2r_1]$ [cf.~Eq.~(\ref{Eq:ropt_alpha_neff})].

For further insight about the dynamics, one can eliminate the cavity-2 from the linearized equation of motion of cavity-1 (see Appendix \ref{App:TwoCav}). Doing so, one can define a total squeezing interaction $\lambda_{1, \mathrm{tot}}$ (analoguous to the DPA, see Eq.~\eqref{HDPA}) which is composed of two contributions: a direct interaction $\lambda_{1}$ coming from the Kerr interaction inside cavity-1, and an induced one $\lambda_{1,\mathrm{ind}}[\omega]$ coming from the interaction with cavity-2; this induced interaction is  frequency dependent. 
In the limit $U_{1,2} \ll \kappa$ ($\kappa_1 = \kappa_2 = \kappa$), these two contributions are ($\Delta_1 = \Delta_2 = \Delta$)
\begin{align}
	\lambda_{1} &= 2U_1 \alpha_1^2, &
	\lambda_{1,\mathrm{ind}} [\omega=0]&\approx 2U_2 \alpha_2^2 \frac{J^2}{\Delta^2 + \kappa^2/4}.
\end{align}
Still in the limit $U_{1,2} \ll \kappa$, we also have
\begin{align}
	\alpha_1^2 \approx \frac{F^2 (\Delta + i\kappa/2)^2}{((\Delta + i\kappa/2)^2 - J^2)^2}, \quad \alpha_2^2 \approx \frac{F^2 J^2}{((\Delta + i\kappa/2)^2 - J^2)^2}.
\end{align} 
In the case where $J$ is the largest scale in the system and where $\Delta \sim \kappa$ (as in Fig.~\ref{Fig:TwoCavities} and Refs~\cite{Liew_Savona_PRL_2010, Bamba_PRA_2011}), the induced squeezing interaction is enhanced compared to the direct contribution as $\lambda_{1,\mathrm{ind}}/\lambda_{1} \sim J^4/\kappa^4$. 
This enhancement results both from the explicit factor of $J^2$ in $\lambda_{1,\mathrm{ind}}$, and from the fact that $\alpha_2/\alpha_1\sim J/\kappa$.
Setting $U_1 = 0$ would actually lead to almost the same results, as pointed out in \cite{Bamba_PRA_2011}.
In short, the second cavity in this system acts to produce an effective squeezing interaction in cavity-1; by tuning $U_2$, the amount of squeezing can be tuned to give an optimal $\g20$ suppression. 
We stress that in general, all that one needs is sufficiently tuned squeezing interaction; having two cavity modes is not necessary.

The almost-complete suppression of $\gone$ found in Ref.~\cite{Bamba_PRA_2011} and shown in Fig.~\ref{Fig:TwoCavities} thus corresponds to having a displaced squeezed state which is infinitesimally close to the vacuum, and where the squeezing is nearly optimally matched to the displacement.  While the large suppression of $\gone$ is interesting, the fact that the state corresponds to almost no photon in cavity-1 (i.e.~ $\bar{n}_{\mathrm{tot},1}$ [cf.~Eq.~(\ref{Eq:ntot})], is of the order of $10^{-7}$) makes the state inconvenient for applications.  While the general optimal-squeezing mechanism allows for $\g20$ suppression at much larger average photon numbers, the required tuning of parameters is hard to achieve in the two cavity setup.  If one simply increases the drive strength in the two-cavity system to increase photon number, $\g20$ is far from optimal  (e.g.~for $F=10\kappa$, $\ntot=0.5$ and $\g20=0.9$ at $U/\kappa=0.006$).
To that extent, the DPA constitutes a simpler and more efficient system to exploit the nonclassical features of amplitude-squeezed Gaussian states \cite{Koashi_PRL_1993, Lu_PRL_2001, Grosse_PRL_2007}.

\section{Conclusion}

In this work, we have discussed how one can achieve classically forbidden values of the normalized intensity fluctuations of a light field ($\g20 < 1$), antibunching ($\g20 < \gt$) and nonclassical non-monotonic behaviors of $\gt$ ($\vert \gt - 1 \vert > \vert \g20 - 1 \vert$) with Gaussian states. The key ingredient is a well-tuned amount of amplitude squeezing. We found that for a fixed average cavity amplitude, there is a minimum possible value of $\g20$ achievable with a Gaussian state; this minimum is attained using a pure state with an optimal amount of amplitude squeezing. This result thus allows one to safely identify non-Gaussian states from a measurement of $\g20$.  We then reviewed how a generic DPA appears to be one of the simplest and most efficient platforms to exploit these nonclassical signatures.  We also demonstrated that it compares favourably to the $\g20$ suppression possible using dissipative squeezing interactions generated via reservoir engineering.  
Finally, we have helped clarify the origin of the so-called unconventional photon blockade predicted in a driven two-cavities setup with weak Kerr nonlinearity, showing that it is a particular realization of this general Gaussian squeezed-state physics. 

\section*{Acknowledgements}

We thank Christoph Marquardt, Florian Marquardt, Cristiano Ciuti and Alexandre Blais for helpful conversations.  This work was supported by NSERC and FQRNT.  ND acknowledges support from CIFAR.

\vspace*{1mm}

\appendix


\section{Optimization of $\g20$ for non-ideal squeezing}\label{App:Opt_g2}

In the case of a finite effective temperature $\neff$, $\g20$ is minimized as follows. First, we choose the right angles such that squeezing is along the displacement (amplitude squeezing) and afterward, we minimize in function of $\ba$ keeping $r$ and $\neff$ constant. It then gives
\begin{equation}
	\aopt[r, \neff] = \sqrt{\frac{(2\neff+1)(e^r \neff + \sinh r)\sinh2r}{e^{-3r}(e^{2r}-(2\neff+1))}}. \label{Eq:alphaOpt_rneff}
\end{equation}
Note that $\aopt[r, \neff]$ is real and positive only if $r \geq \ln\left( \sqrt{2\neff+1}\right)$. For squeezing below this minimal value, it is possible to show that $\g20 > 1$.

We then minimize $\g20$ (with $\ba = \aopt$) in function of $r$ keeping $\neff$ constant, which leads to $\sinh^2\ropt=\neff$ (Eq.~\eqref{Eq:ropt})
and 
\begin{widetext}
\begin{gather}
	\g20 \vert_{\mathrm{opt}} = 1 - \frac{1}{1+24\neff + 2\neff^2(11 + 8\neff(2 + \neff)) + 8(1 + 6\neff + 12\neff^2 + 8\neff^3)\sqrt{\neff(\neff + 1)}}.
\end{gather}
\end{widetext}


\section{Degenerate parametric amplifiers}\label{App:DPA}

\subsection{Calculation of $g^{(2)}(t)$}

The Langevin equation induced by the general Hamiltonian Eq.~\eqref{HDPA} of the DPA reads
\begin{equation} \label{Eq:EMO_DPA}
\partial_t\hat{d}(t) = -\lambda e^{i\theta} \hat{d}^\dag(t)-\tfrac{1}{2}\kappa\,\hat{d}(t)-\sqrt{\kappa}\,\hat{d}_\mathrm{in}(t),
\end{equation} 
where we define $\hat{d}=\hat{a}-\alpha$, $\alpha=\langle\hat{a}\rangle$ 
and we note $\kappa$ the damping rate of the resonator.
The operator $\hat{d}_\mathrm{in}$ is the input noise corresponding to vacuum, that is $\langle\hat{d}_\mathrm{in}(t)\rangle=0$ and the only non-vanishing correlation function is $\langle\hat{d}_\mathrm{in}(t)\hat{d}^\dag_\mathrm{in}(0)\rangle=\delta(t)$.
The Langevin equation is integrated into
\begin{multline}
\hat{d}(t) = -\sqrt{\kappa}\,\int\limits_{-\infty}^t\mathrm{d}t'
e^{-\kappa(t-t')/2}\left\{\cosh[\lambda(t-t')]\hat{d}_\mathrm{in}(t')\right.\\\left.-e^{i\theta}\sinh[\lambda(t-t')]\hat{d}^\dag_\mathrm{in}(t')\right\}.
\end{multline} 
The two-time correlations are then equal to
\begin{align}
\langle\hat{d}^\dag(t)\hat{d}(0)\rangle&=
\frac{\lambda\,e^{-\kappa t/2}}{\kappa^2-4\lambda^2}\left[2\lambda\cosh\lambda t+\kappa\sinh\lambda t\right],\\
\langle\hat{d}(t)\hat{d}(0)\rangle&=
-\frac{\lambda\,e^{-\kappa t/2}\,e^{i\theta}}{\kappa^2-4\lambda^2}\left[2\lambda\sinh\lambda t+\kappa\cosh\lambda t\right].
\end{align}
Noting 
$n(t)=\langle\hat{d}^\dag(t)\hat{d}(0)\rangle$,
$s(t)=|\langle\hat{d}(t)\hat{d}(0)\rangle|$,
$\ba=|\alpha|$,
$\varphi=\arg\alpha$,
$g^{(2)}(t)$ is found from
\begin{align}
g^{(2)}(t)&=1+\frac{2\ba^2[n(t)-\cos(\theta-2\varphi)s(t)]+n^2(t)+s^2(t)}{(\ba^2+n(0))^2}.
\label{g2tgeneral}
\end{align}

\subsection{Thermal squeezed state parameters}

At coinciding times, the two-time correlations give
\begin{align}
\langle\hat{d}^\dag\hat{d}\rangle&=
\frac{2\lambda^2}{\kappa^2-4\lambda^2},&
\langle\hat{d}^2\rangle&=
-\frac{\kappa\lambda\,e^{i\theta}}{\kappa^2-4\lambda^2}.
\end{align}
Comparing these results to the case of a squeezed thermal state,
$\mathrm{Tr}[\hat{d}^\dag\hat{d}\,\hat{\rho}_{\neff,\xi}]=\neff\cosh2r+\sinh^2r$
and
$\mathrm{Tr}[\hat{d}^2\,\hat{\rho}_{\neff,\xi}]=-(\neff+\frac{1}{2})\sinh2r\,e^{i\theta}$,
we get
\begin{align}
\tanh2r&=\frac{2\lambda}{\kappa},&
\neff&=\sinh^2r.
\end{align}

\subsection{Output field} \label{App:OutputField}

From input-output formalism~\cite{Clerk_RMP_2010, GardinerZollerBook}, the output field $\hat{a}_\mathrm{out}=\hat{d}_\mathrm{out}+\alpha_\mathrm{out}$ is given by
\begin{equation}
\hat{d}_\mathrm{out}(t)=\sqrt{\kappa}\hat{d}(t)+\hat{d}_\mathrm{in}(t),
\end{equation}
giving rise, for $t\geq0$, to the output correlations
\begin{subequations}
\begin{align}
\langle\hat{d}_\mathrm{out}^\dag(t)\hat{d}_\mathrm{out}(0)\rangle = &
\kappa\langle\hat{d}^\dag(t)\hat{d}(0)\rangle
+\sqrt{\kappa}\langle\hat{d}_\mathrm{in}^\dag(t)\hat{d}(0)\rangle \label{expendDd} \\ 
&+\sqrt{\kappa}\langle\hat{d}^\dag(t)\hat{d}_\mathrm{in}(0)\rangle
+\langle\hat{d}_\mathrm{in}^\dag(t)\hat{d}_\mathrm{in}(0)\rangle
\nonumber \\
= &\kappa\langle\hat{d}^\dag(t)\hat{d}(0)\rangle, \\
\langle\hat{d}_\mathrm{out}(t)\hat{d}_\mathrm{out}(0)\rangle =&
\kappa\langle\hat{d}(t)\hat{d}(0)\rangle
+\sqrt{\kappa}\langle\hat{d}_\mathrm{in}(t)\hat{d}(0)\rangle  \label{expenddd} \\
& +\sqrt{\kappa}\langle\hat{d}(t)\hat{d}_\mathrm{in}(0)\rangle
+\langle\hat{d}_\mathrm{in}(t)\hat{d}_\mathrm{in}(0)\rangle
\nonumber \\
= &\kappa\langle\hat{d}(t)\hat{d}(0)\rangle.
\end{align}
\end{subequations}
In Eqs.~\eqref{expendDd} and~\eqref{expenddd}, the second term is eliminated due to causality: the input fluctuations at time $t>0$ cannot affect the intracavity field at time $t=0$. The third and fourth terms also vanish since the incoming noise is vacuum noise.
Consequently, $n_\mathrm{out}(t)=\kappa n(t)$ and $s_\mathrm{out}(t)=\kappa s(t)$; note also that $[\hd_{\mathrm{out}}(t),\hd_{\mathrm{out}}^\dag(0)] = \delta(t)$ and $[\hd_{\mathrm{out}}(t),\hd_{\mathrm{out}}(0)]=0$, as it should.

Concerning the average value of the output field, $\alpha_\mathrm{out}=\langle\hat{a}_\mathrm{out}\rangle$, we have
\begin{equation}
\alpha_\mathrm{out}=\sqrt{\kappa}\alpha+\alpha_\mathrm{in}.
\end{equation}
From the general expression Eq.~\eqref{g2tgeneral} of $g^{(2)}(t)$, we see that $g^{(2)}_\mathrm{out}(t)$ is equal to $g^{(2)}(t)$ with a renormalized displacement: 
$\alpha\to\alpha+\alpha_\mathrm{in}/\sqrt{\kappa}$.


\section{Squeezed environment} \label{App:SqueezedEnv}

\subsection{Calculation of $\gt$}

As discussed in the main text, we model QBE by considering a two-sided linear cavity coupled on one side to an environment in a vacuum squeezed state and driven on the other side, both on resonance with the cavity frequency.
The damping rate of the former side is $\Gamma_{\rm QBE}$ and the one of the latter is $\kappa_{\rm ext}$, we note $\kappa_\mathrm{QBE}=\Gamma_{\rm QBE}+\kappa_{\rm ext}$.
The operator of the squeezed vacuum is $\hat{b}_\mathrm{in}$
and the operator of the drive is $\hat{c}_\mathrm{in}$.
In an interaction picture at the cavity resonance frequency, the Langevin equation of the intracavity field is
\begin{equation}
\partial_t\hat{a}(t)=-\tfrac{1}{2}\kappa_\mathrm{QBE}\,\hat{a}(t)-\sqrt{\Gamma_{\rm QBE}}\,\hat{b}_\mathrm{in}(t)-\sqrt{\kappa_{\rm ext}}\,\hat{c}_\mathrm{in}(t),
\end{equation}
where the input noise corresponds to a displaced vacuum squeezed state,
\begin{align}
\langle \hat{b}_\mathrm{in}^\dag(t) \hat{b}_\mathrm{in}(0) \rangle&=\sinh^2r_\mathrm{QBE}\,\delta(t),\\
\langle \hat{b}_\mathrm{in}(t) \hat{b}_\mathrm{in}(0) \rangle&=-\tfrac{1}{2}\sinh2r_\mathrm{QBE}\,e^{i\theta}\,\delta(t),\\
\langle \hat{c}_\mathrm{in}(t) \rangle&=\alpha_\mathrm{in}=-\frac{\kappa_\mathrm{QBE}}{2\sqrt{\kappa_{\rm ext}}}\alpha,
\end{align}
where $\hat{c}_\mathrm{in}=\alpha_\mathrm{in}+\hat{d}_\mathrm{in}$.
The operator $\hat{d}_\mathrm{in}$ describes the same vacuum noise as in Section~\ref{App:DPA}.
The intracavity quantum fluctuations $\hat{d}=\hat{a}-\alpha$ are then equal to
\begin{align}
\hat{d}(t)=&-\sqrt{\Gamma_{\rm QBE}}\int_{-\infty}^t\mathrm{d}t'\,e^{-\frac{1}{2}\kappa_\mathrm{QBE}(t-t')}\hat{b}_\mathrm{in}(t')\nonumber\\
&-\sqrt{\kappa_{\rm ext}}\int_{-\infty}^t\mathrm{d}t'\,e^{-\frac{1}{2}\kappa_\mathrm{QBE}(t-t')}\hat{d}_\mathrm{in}(t').
\end{align}
The correlations of the fluctuations are
\begin{align}
\langle\hat{d}^\dag(t)\hat{d}(0)\rangle&=\eta\sinh^2r_\mathrm{QBE}\,e^{-\frac{1}{2}\kappa_\mathrm{QBE} t}, \\
\langle\hat{d}(t)\hat{d}(0)\rangle&=-\tfrac{1}{2}\eta\sinh2r_\mathrm{QBE}e^{i\theta}\,e^{-\frac{1}{2}\kappa_\mathrm{QBE} t},
\end{align}
where we note $\eta=\Gamma_{\rm QBE}/\kappa_\mathrm{QBE}$.
The two-time second-order correlations read 
\begin{widetext}
\begin{equation} \label{Eq:g2QBE}
g^{(2)}(t)=1-\frac{\frac{1}{\eta}|\alpha|^2(1-e^{-2r_\mathrm{QBE}})\,e^{-\frac{1}{2}\kappa_\mathrm{QBE} t}-\sinh^2r_\mathrm{QBE}\cosh2r_\mathrm{QBE}\,e^{-\kappa_\mathrm{QBE} t}}{(\frac{1}{\eta}|\alpha|^2+\sinh^2r_\mathrm{QBE})^2}.
\end{equation}
\end{widetext}

Note that the output field on the driven side is obtained from
\begin{align}
\hat{d}_\mathrm{out}&=\sqrt{\kappa_{\rm ext}}\hat{d}+\hat{d}_\mathrm{in},&
\alpha_\mathrm{out}&=\sqrt{\kappa_{\rm ext}}\alpha+\alpha_\mathrm{in},
\end{align}
with the output correlations
\begin{align}
\langle\hat{d}_\mathrm{out}^\dag(t)\hat{d}_\mathrm{out}(0)\rangle
&=\kappa_{\rm ext} \langle\hat{d}^\dag(t)\hat{d}(0)\rangle,\\
\langle\hat{d}_\mathrm{out}(t)\hat{d}_\mathrm{out}(0)\rangle
&=\kappa_{\rm ext} \langle\hat{d}(t)\hat{d}(0)\rangle,
\end{align}
for $t\geq0$.
The output intensity correlations are then given by Eq.~\eqref{Eq:g2QBE} by replacing $\alpha \rightarrow \alpha_{\mathrm{out}}/\sqrt{\kappa_{\rm ext}}$.

\subsection{Comparison with the DPA}
\label{App:QBEcomp}

The intensity correlations obtained with QBE are compared to the case of the DPA by setting the same total damping rate $\kappa$ and the same intracavity squeeze parameter $r$ (amplitude squeezing),
which is achieved by setting
\begin{equation}
e^{2r_\mathrm{QBE}}=\frac{2\lambda(1-\eta)+\sqrt{4\lambda^2(1-2\eta)+\eta^2\kappa^2}}{\eta(\kappa-2\lambda)}.
\end{equation}
The purity $P=1/(1+2\bar{n}_\mathrm{eff})$ of the two systems is then
\begin{equation}
P_\mathrm{DPA}=\sqrt{1-4\lambda^2/\kappa^2},
\end{equation}
and
\begin{multline}
P_\mathrm{QBE}=\Big[1+\frac{2(1-\eta)}{\kappa^2-4\lambda^2}\Big(4\lambda^2-\eta^2\kappa^2\\
+\sqrt{4\lambda^2(1-2\eta)+\eta^2\kappa^2}\Big)\Big]^{-1/2}.
\end{multline}
The purity from QBE is higher than the state of the DPA for $\eta>1/2$
and always tends to unity as $\eta\to1$ for any value of $\lambda/\kappa$.

For a given intracavity squeezing, the optimal equal-time intensity correlation is equal to
\begin{equation}
g^{(2)}_{\substack{\mathrm{opt}\\\mathrm{QBE}}}=
1-\frac{2}{e^{4r_\mathrm{QBE}}+2e^{2r_\mathrm{QBE}}-1},
\end{equation}
obtained at the optimal displacement
\begin{equation}
\bar{\alpha}_{\substack{\mathrm{opt}\\\mathrm{QBE}}}^2=
\tfrac{1}{4}\eta(e^{4r_\mathrm{QBE}}-1).
\end{equation}
The optimal $\g20$ is lower with QBE for $\eta>1/2$.
The purity and optimal $\g20$ coincide at $\eta=1/2$, i.e. for $\Gamma_{\rm QBE}=\kappa_{\rm ext}$.

\section{Solution of the quantum dynamics of the two coupled cavities} \label{App:TwoCav}

\subsection{Linearized equations of motion}

In this section, we give details about the linearization of the equations of motion of the two coupled cavities (see Eq.~\eqref{Eq:TwoCavHamiltonian}) and the technique used to solve them. 

As is standard for optical cavities, we treat the environment as Markovian dissipative baths with zero effective temperature. Using input-output formalism \cite{Clerk_RMP_2010,GardinerZollerBook}, we write the full nonlinear equations of motion as:
\begin{align}
	& \partial_t\ha_k = -i \left[ \ha_k, \hat{H}\right] - \frac{\kappa_k}{2} \ha - \sqrt{\kappa_k}\hat{\xi}_k, \nonumber \\ 
	& =  \left[ i\Delta_k  - \frac{\kappa_k}{2} \right]\ha_k - 2i U_k \had_k\ha_k\ha_k - i J \ha_{l} - \sqrt{\kappa_k} \hat{\xi}_k - i F \delta_{k,1}.
\end{align}
Here, $\kappa_k$ is the damping rate of cavity $k$, $\hat{\xi}_k$ is the annihilation operator of the corresponding incoming vacuum noise, $l = 1$ ($l = 2$) if $k= 2$ ($k=1$) and $\delta_{k,1}$ is the Kronecker function. The only non-vanishing correlation function of the vacuum noise is:
\begin{equation}
	\langle \hat{\xi}_k(t)\hat{\xi}_k^{\dag}(t') \rangle = \delta(t-t').
\end{equation}

As presented in the main text, we want to show that the corresponding linearized theory is enough to recover the results predicted in \cite{Bamba_PRA_2011}. To do this, we first displace the field operators by their steady-state mean values, i.e.~$\ha_k \rightarrow \alpha_k + \hd_k$ with $\langle \hd_k \rangle = 0$.
Doing so, one gets
\begin{subequations}
\begin{align}
& \partial_t \hd_k = \left[ i(\Delta_k - 4U_k \vert\alpha_k\vert^2)  - \frac{\kappa_k}{2} \right]\hd_k - 2i U_k \alpha_k^2 \hdd_k  \nonumber \\
	& - 2i U_k(\hdd_k\hd_k + \alpha_k^* \hd_k + 2\alpha_k \hdd_k )\hd_k - i J \hd_{l} - \sqrt{\kappa_k} \hat{\xi}_k, \label{Eq:Qu_Nl_EOM} \\ 
	& \partial_t \alpha_k = \left[ i\Delta_k  - \frac{\kappa_k}{2} \right]\alpha_k - 2i U_k\vert \alpha_k \vert^2 \alpha_k - i J \alpha_{l} - i F \delta_{k,1}. \label{Eq:Cl_Nl_EOM}
\end{align}
\end{subequations}
From there, we drop the nonlinear terms in the quantum equations of motion Eq.~(\ref{Eq:Qu_Nl_EOM}). It then leads to
\begin{subequations} \label{Eq:EOM}
\begin{align}
	\partial_t \hd_k  = & \left[ i(\Delta_k - 4U_k \vert\alpha_k\vert^2)  - \frac{\kappa_k}{2} \right]\hd_k \nonumber\\&- 2i U_k \alpha_k^2 \hdd_k - i J \hd_{l} - \sqrt{\kappa_k} \hat{\xi}_k, \label{Eq:Qu_EOM} \\ 
	0 = & \left[ i\Delta_k  - \frac{\kappa_k}{2} \right]\alpha_k - 2i U_k\vert \alpha_k \vert^2 \alpha_k - i J \alpha_{l} - i F \delta_{k,1}. \label{Eq:Cl_EOM}
\end{align}
\end{subequations}
The coupled nonlinear classical equations are straightforward to solve, but the solutions are too cumbersome to be shown here. In order to solve the quantum counter-part of the Heisenberg-Langevin equations, one can go in the frequency domain, defining (same for $\hat{\xi}_k[\omega]$)
\begin{equation}
	\hd_k [\omega] \equiv \int_{-\infty}^{\infty} \hd_k(t) e^{i\omega t}, \qquad 
	\hdd_k [\omega] \equiv \int_{-\infty}^{\infty} \hdd_k(t) e^{i\omega t}.
\end{equation}
In the frequency domain, Eqs.~(\ref{Eq:Qu_EOM}) becomes 
\begin{equation}
\vec{\hd}[\omega] = -\sqrt{\kappa} \chi[\omega] \vec{\hat{\xi}}[\omega],
\end{equation}
with $\vec{\hd}[\omega] = \left[\hd_1[\omega], \hd_2[\omega], \hdd_1[\omega], \hdd_2[\omega]\right]^{\intercal}$, $\vec{\hat{\xi}}[\omega] = \left[\hat{\xi}_1[\omega], \hat{\xi}_2[\omega], \hat{\xi}^{\dag}_1[\omega], \hat{\xi}^{\dag}_2[\omega]\right]^{\intercal}$ and the susceptibility 
\begin{widetext}
\begin{equation}
	\chi[\omega] \equiv \begin{bmatrix}
	\kappa_1/2 - i(\Delta_1 + \omega) & i J & 2iU_1\alpha_1^2 & 0\\
	i J & \kappa_2/2 - i(\Delta_2 + \omega) & 0 & 2iU_2\alpha_2^2 \\
	-2i U_1 (\alpha_1^*)^2 & 0 & \kappa_1/2 - i(-\Delta_1 + \omega) & -i J \\
	0 & -2iU_2(\alpha_2^*)^2 & -i J	& \kappa_2/2 - i(-\Delta_2 + \omega)
	\end{bmatrix}^{-1}.
\end{equation}
\end{widetext}

In terms of the $\chi[\omega]$ matrix, one has (see Eqs.~\eqref{Eq:ns})
\begin{align}
	n_k & \equiv \langle \hdd_k(t=0) \hd_k(t=0) \rangle \nonumber \\
	& = \int_{-\infty}^{\infty} \frac{d\omega}{2\pi} \left( \vert\chi_{k,3}[\omega]\vert^2 + \vert\chi_{k,4}[\omega]\vert^2 \right),
\end{align}
and
\begin{align}
	s_k & \equiv \vert \langle \hd_k(t=0) \hd_k(t=0) \rangle \vert \nonumber \\
	&= \left\vert \int_{-\infty}^{\infty} \frac{d\omega}{2\pi} \left( \chi_{k+2,3}[\omega]^* \chi_{k,3}[\omega] + \chi_{k+2,4}[\omega]^* \chi_{k,4}[\omega]  \right) \right\vert.
\end{align}

Finally, using Eq.~\eqref{Eq:Qu_EOM}, one can express $\hdd_2[\omega]$ only in terms of $\hd_1[\omega]$, $\hdd_1[\omega]$ and noise operators and then formally eliminate cavity-2 of the equation of motion for cavity-1. It leads to (the frequency dependence of the operators is implicit for clarity)
\begin{align} \label{Eq:EOMCav1}
	\hd_1 = & \left( i(\omega + \Delta_1) - \kappa_1/2 - i J^2 G^R_2 [\omega] \right)\hd_1 \nonumber \\
	& -i \left( 2U_1 \alpha_1^2 - \frac{2 J^2 U_2 \alpha_2^2}{\omega - \Delta_2 + i \kappa_2/2} G^R_2[\omega] \right) \hdd_1 \\
	& -\sqrt{\kappa_2} J G^R_2[\omega]\left( \hat{\xi}_2 + \frac{2U_2 \alpha_2^2}{\omega - \Delta_2 + i \kappa_2/2} \hat{\xi}^{\dag}_2 \right) - \sqrt{\kappa_1}\hat{\xi}_1, \nonumber
\end{align}
with 
\begin{equation}
	G^R_2 [\omega ] \equiv \frac{1}{\omega + \Delta_2 + i\kappa_2/2 + \frac{4U_2^2 \vert \alpha_2 \vert^4}{\omega - \Delta_2 + i \kappa_2 /2}}.
\end{equation}
From Eq.~\eqref{Eq:EOMCav1}, one sees that one of the net effects of cavity-2 is to generate an additional parametric interaction into cavity-1. If we compare with the DPA (see Eq.~\eqref{Eq:EMO_DPA}), one can define a total parametric interaction strength $\lambda_{1\mathrm{tot}}$ as
\begin{multline} \label{Eq:lambda1}
	\lambda_{1\mathrm{tot}} \equiv 2U_1 \alpha_1^2 \\ - \frac{2 J^2 U_2 \alpha_2^2}{(\omega - \Delta_2 + i \kappa_2/2)(\omega + \Delta_2 + i \kappa_2/2) - 4U_2^2 \vert \alpha_2 \vert^4}.
\end{multline}
As discussed in the main text, for the parameters used in Fig.~\ref{Fig:TwoCavities}, the main source of squeezing comes from cavity-2.

\subsection{Quantum master equation} \label{App:QME}

The quantum dynamics of the two coupled cavities can be solved numerically by finding the steady state of the density matrix $\hat{\rho}$ of the system.
The time evolution of the density matrix is governed by the Hamiltonian $\hat{H}$ of Eq.~\eqref{Eq:TwoCavHamiltonian} and the Lindbladian $\hat{L}$ of the Markovian environments at zero temperature,
\begin{equation}
\partial_t\hat{\rho}=-i[\hat{H},\hat{\rho}]+\hat{L}\hat{\rho}.
\label{EqQME}
\end{equation}
The Lindbladian
\begin{equation}
\hat{L}=\kappa_1\hat{D}[\hat{a}_1]+\kappa_2\hat{D}[\hat{a}_2],
\end{equation}
is expressed in terms of the damping rates $\kappa_{1,2}$ and the dissipator superoperator~\cite{GardinerZollerBook},
\begin{equation}
\hat{D}[\hat{a}]\cdot=\hat{a}\cdot\hat{a}^\dag-\tfrac{1}{2}\{\hat{a}^\dag\hat{a},\cdot\}.
\end{equation}

\begin{figure}[t]
  \centering
  \includegraphics[width=\columnwidth]{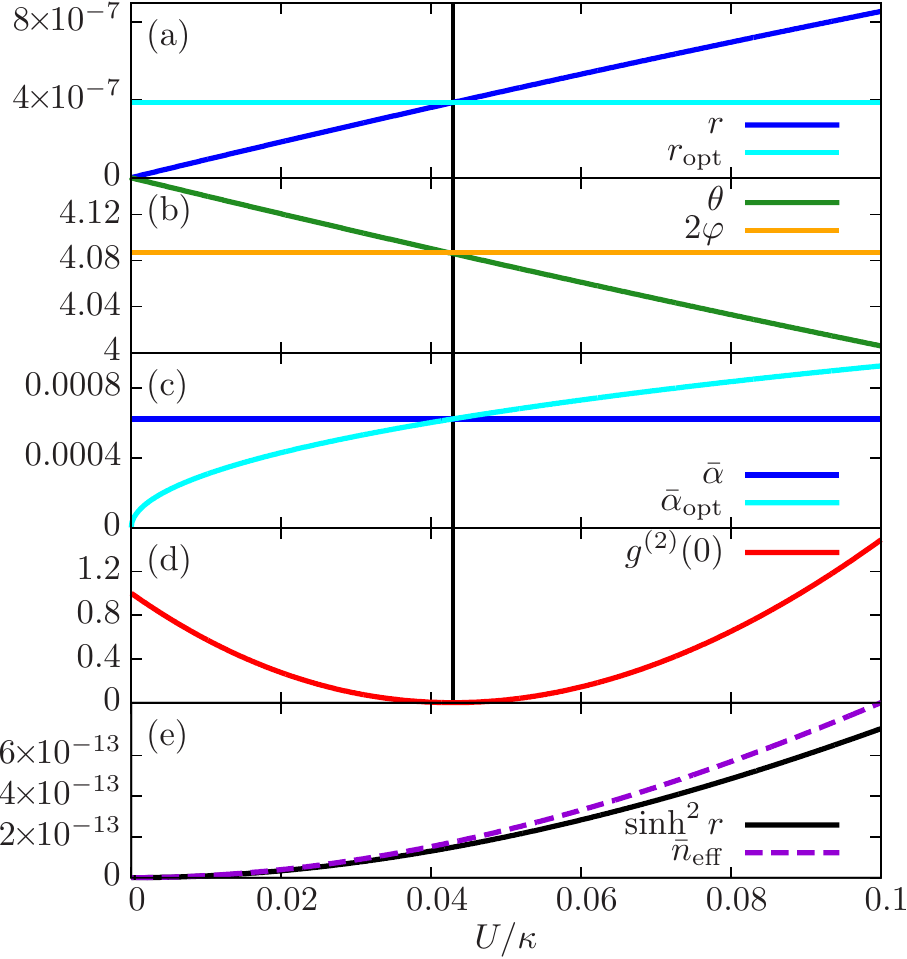}
  \caption{Steady-state solution of the quantum master equation Eq.~\eqref{EqQME} for cavity-1: 
(a) Squeeze parameter $r_1$ and optimal value $r_{\mathrm{opt}1}$;
(b) Squeezing angle $\theta_1$ compared to twice the displacement argument $\varphi_1$;
(c) Displacement $\ba_1$ and optimal value $\ba_{\mathrm{opt}1}$;
(d) Intensity correlation at equal times $g^{(2)}_1(0)$;
(e) Effective photon number $\bar{n}_{\mathrm{eff}1}$ and $\sinh^2r_1$.
The parameters are plotted against $U\equiv U_1=U_2$ for $\kappa_1=\kappa_2\equiv\kappa$ and the same parameters as in Fig.~\ref{Fig:TwoCavities}.
The vertical line is placed at the minimal value of the intensity correlation, and coincides with $r=r_\mathrm{opt}$, $\theta=2\varphi$ and $\ba=\aopt$.}
\label{Fig:Appendix}
\end{figure}

The steady-state field $\langle\hat{a}_1\rangle$, photon number $\langle\hat{a}_1^\dag\hat{a}_1\rangle$, squeezing $\langle\hat{a}_1^2\rangle$, equal-time intensity correlation $g_1^{(2)}(0)$, as well as the parameters of the corresponding displaced squeezed thermal state for the cavity-1 are plotted in Fig.~\ref{Fig:Appendix} as a function of the interaction strength $U$.
The results are in very good agreement with the linearized theory, especially around the minimum of the intensity correlation.
As predicted from the linearized theory, this minimum coincides with the optimal condition for the squeeze parameter, $r=r_\mathrm{opt}\approx\bar{\alpha}^2/(1+2\bar{\alpha}^2)$ (for $r\ll1$) and the displacement, $\ba=\aopt$. Moreover, considering the full nonlinear dynamics, we see that $\theta$ depends on $U$ and that the minimum of $\g20$ coincides with amplitude squeezing, i.e.~$\theta=2\varphi$.

For large tunnel coupling and drive, it is more convenient to work with the normal modes $\hat{b}_1$ and $\hat{b}_2$ that diagonalize the Hamiltonian without nonlinearity.
These normal modes are defined so as to eliminate the linear and tunneling terms
\begin{equation}
\begin{pmatrix}\hat{a}_1\\\hat{a}_2\end{pmatrix}
=
\begin{pmatrix}\cos\phi&\sin\phi\\-\sin\phi&\cos\phi\end{pmatrix}
\begin{pmatrix}\hat{b}_1\\\hat{b}_2\end{pmatrix}
+
\begin{pmatrix}\alpha_1\\\alpha_2\end{pmatrix},
\end{equation}
with
\begin{align}
\tan2\phi&=\frac{2J}{\Delta_1-\Delta_2}, \\
\alpha_1&=\frac{(\Delta_2+\frac{1}{2}i\kappa_2)F}{(\Delta_1+\frac{1}{2}i\kappa_1)(\Delta_2+\frac{1}{2}i\kappa_2)-J^2}, \\
\alpha_2&=\frac{JF}{(\Delta_1+\frac{1}{2}i\kappa_1)(\Delta_2+\frac{1}{2}i\kappa_2)-J^2}.
\end{align}
The Hamiltonian then reads
\begin{align}
\hat{H}= & -(\Delta_1\cos^2\phi+\Delta_2\sin^2\phi+\tfrac{1}{2}J\sin2\phi)\hat{b}_1^\dag\hat{b}_1 \nonumber \\
& -(\Delta_1\sin^2\phi+\Delta_2\cos^2\phi-\tfrac{1}{2}J\sin2\phi)\hat{b}_2^\dag\hat{b}_2 \nonumber \\
&+\sum_{k=1}^2U_k \had_k\had_k\ha_k\ha_k.
\end{align}
One has finally to express the Kerr nonlinearities as well as the Lindbladian in terms of the $\hat{b}_k$.

\bibliographystyle{apsrev}
\bibliography{MALPapersRefs}

\end{document}